\pdfoutput=1
\RequirePackage{ifpdf}
\ifpdf 
\documentclass[pdftex]{sigma}
\else
\documentclass{sigma}
\fi

\numberwithin{equation}{section}

\usepackage{units}
\usepackage{multirow}
\usepackage{upgreek}
\usepackage{xargs}[2008/03/08]

\global\long\def\set#1#2{\left\{ #1\, |\, #2\right\} }%

\global\long\def\inv{^{\,\textrm{-}1}}%

\global\long\def\sn#1{\mathbb{S}_{#1}}%

\global\long\def\FA#1{\vert#1\vert}%

\global\long\def\FC#1{#1^{{\scriptscriptstyle \mathtt{o}}}}%

\newcommandx\exi[3][usedefault, addprefix=\global, 1=, 2=]{\mathtt{e}^{\mathtt{{\scriptscriptstyle \textrm{#1}}}\frac{#2\pi\mathtt{i}}{#3}}}%

\global\long\def\stab#1#2{#1_{#2}}%

\global\long\def\cent#1#2{\mathtt{C}_{#1} \left(#2\right)}%

\global\long\def\chb#1{\xi_{#1}}%

\global\long\def\chg{\mathfrak{g}}%

\global\long\def\qd#1{\mathtt{d}_{#1}}%
\global\long\def\cw#1{\mathtt{h}_{#1}}%

\global\long\def\tcl{\boldsymbol{\mathfrak{1}}}%

\global\long\def\rami#1{\mathtt{e}_{#1}}%

\global\long\def\el#1{\FA{\FC{#1}}}%

\global\long\def\cs#1{\left\Vert #1\right\Vert }%

\global\long\def\irr#1{\mathtt{Irr} \left(#1\right)}%

\global\long\def\to{\cs{\chg}}%

\global\long\def\v{\mathtt{{\scriptstyle 0}}}%

\global\long\def\fm{\mathtt{N}}%

\global\long\def\spr{\textrm{spread}}%

\global\long\def\bfm{\mathsf{N}}%

\begin{document}

\allowdisplaybreaks

\newcommand{\arXivNumber}{1810.00145}

\renewcommand{\thefootnote}{}

\renewcommand{\PaperNumber}{027}

\FirstPageHeading

\ShortArticleName{A Short Guide to Orbifold Deconstruction}

\ArticleName{A Short Guide to Orbifold Deconstruction\footnote{This paper is a~contribution to the Special Issue on Moonshine and String Theory. The full collection is available at \href{https://www.emis.de/journals/SIGMA/moonshine.html}{https://www.emis.de/journals/SIGMA/moonshine.html}}}

\Author{Peter BANTAY}

\AuthorNameForHeading{P.~Bantay}

\Address{Institute for Theoretical Physics, E\"otv\"os L\'or\'and University, H-1117 Budapest,\\
 P\'azm\'any P\'eter s.~1/A, Hungary}
\Email{\href{mailto:bantay@poe.elte.hu}{bantay@poe.elte.hu}}

\ArticleDates{Received September 28, 2018, in final form March 27, 2019; Published online April 09, 2019}

\Abstract{We study the problem of orbifold deconstruction, i.e., the process of recognizing, using only readily available information, whether a given conformal model can be realized as an orbifold, and the identification of the twist group and the original conformal model.}


\Classification{11F99; 13C05}

\renewcommand{\thefootnote}{\arabic{footnote}}
\setcounter{footnote}{0}

\section{Introduction}

Orbifolding \cite{Dixon_orbifoldCFT, Dixon_orbifolds1,Dixon_orbifolds2}, i.e., the gauging of (discrete) symmetries of a conformal model, is one of the most important procedures for obtaining new models from known ones, especially since it respects such important properties as the central charge, unitarity and rationality\footnote{It has even been argued that all rational conformal models may be obtained either as GKO cosets or as orbifolds thereof.}. Unfortunately, the precise determination of the properties of the orbifold could be quite involved in general, and effective techniques are only known for special types, like holomorphic \cite{Bantay1990a,Bantay1991,DPR, DV3} or permutation orbifolds \cite{Bantay1998a,Bantay2002,BDM,Borisov-Halpern-Schweigert,FKS, Klemm-Schmidt}.

The present note addresses the inverse problem: instead of trying to determine the structure of an orbifold from the knowledge of the original model and the action of the twist group, we ask for effective procedures to recognize whether a conformal model could be obtained as an orbifold, and if so, to identify the original model together with the relevant twist group. That such procedures could exist is actually not that surprising, as simple current extensions~\cite{Bantay1998, Fuchs1996a} are nothing but an example of this for Abelian twist groups. But the question is how the corresponding results could be extended to more general orbifolds, and what kind of new structures emerge in the process. We should stress that our approach in the present note is computational: we aim to identify in a down-to-earth manner the relevant conformal models with the help of some simple, mostly numerical data (like conformal weights, quantum dimensions and fusion rules of the primary fields) \cite{DiFrancesco-Mathieu-Senechal}, providing explicit expressions for the characteristic quantities of the deconstructed model. We have no doubt that the whole process could be described elegantly in more abstract terms \cite{FFRS, Kirillov2004}, but the relevant techniques seem (at least to us) less amenable to practical computations.

In the next section we'll summarize those general features of orbifold constructions that will form the basis of the deconstruction procedure. Then we'll turn to the description of the procedure itself, while the last section provides an outlook on questions not addressed in this note. We stress that our exposition is informal, presenting only the most important ideas and their mutual interrelations, but omitting formal arguments. Most of these arguments can be supplied readily, while some of them seem to require more effort, but with hindsight one could argue that a strong supporting evidence of the ideas to be presented is that they lead to a~coherent algorithmic procedure giving meaningful answers that agree with the correct ones in all known cases which could be checked by alternative means. Should some of our assertions prove to be wrong, one would expect that the whole procedure would lead to completely meaningless results in most cases. And it goes without saying that in case of Abelian orbifolds we just get back a~suitable simple current extension~\cite{Fuchs1996a}.

\section{Generalities on orbifolds}\label{sec:Generalities-on-orbifolds}

Consider a (unitary) conformal model whose chiral algebra ${\mathbb V}$ is a well behaved (rational, $C_{2}$-cofinite, etc.) vertex operator algebra \cite{FLM1,Kac,Lepowsky-Li,Zhu1996}. Given a subgroup $G < \operatorname{Aut}({\mathbb V})$ of automorphisms of ${\mathbb V}$, the $G$-orbifold \cite{DV3, Dixon_orbifoldCFT} is obtained by identifying the states that can be transformed into each other by an element of $G$. The Hilbert space of the orbifold is a direct sum of twisted modules of the chiral algebra ${\mathbb V}$ (these consist of those operators which are local with respect to ${\mathbb V}$ only up to an element of the twist group $G$), while its chiral algebra is the fixed-point subalgebra ${\mathbb V}^{G}=\set{v \in {\mathbb V}}{gv=v\textrm{ for all }g \in G}$.

There is a natural action of $G$ on the set of all twisted modules under which an element $h \in G$ sends a $g$-twisted module $M$ to a $hgh\inv$-twisted module $h (M)$. This shows that the set of all $g$-twisted modules, as $g$ runs over a given conjugacy class of the twist group, is $G$-stable under this action, leading to a partition of the orbifold's Hilbert space into sectors labeled by the conjugacy classes of $G$, each of which usually splits into several $G$-orbits. Since the twisted modules inside a given $G$-orbit are all related by the action of an automorphism of ${\mathbb V}$, it follows that they have pretty similar properties (while still not being isomorphic), e.g., their conformal weights $\cw M$, quantum dimensions $\qd M$ and trace functions~\cite{DLM,Zhu1996}
\begin{gather*}
Z_{M} (\tau)=\operatorname{Tr}_{M} \big\{ \mathtt{e}^{2\pi\mathtt{i}\tau(L_{0}-\nicefrac{c}{24})}\big\}
\end{gather*}
all coincide. This means that it is enough to know these data for just one representative module~$M$ from the orbit.

The stabilizer $\stab GM=\set{h \in G}{h (M) \cong M}$ of a $g$-twisted module $M$ is the subgroup of those elements $h \in G$ for which $h (M)$ is isomorphic to~$M$; clearly, it is a subgroup of the centralizer $\cent Gg=\set{h \in G}{gh=hg}$, and the stabilizers of different modules belonging to the same $G$-orbit are all conjugate to each other. Note that the length of the orbit of~$M$, i.e., the number of different modules contained in it, is just the index $[G : G_{M}]$ in $G$ of its stabilizer. By the previous reasoning, we have on each twisted module $M$ an action of its stabilizer $\stab GM$, which is usually not a linear representation, but only a projective one, with associated $2$-cocycle $\vartheta_{M} \in Z^{2}  (G_{M},\mathbb{C} )$. This projective representation decomposes into homogeneous components corresponding to the irreducible projective representations $\chb p \in \irr{G_{M}\vert\vartheta_{M}}$ of the stabilizer, and each such homogeneous component corresponds to a primary field of the orbifold.

\looseness=1 To recapitulate, twisted modules are organized into sectors labeled by conjugacy classes of the twist group $G$, and each such sector splits into orbits of $G$, with each orbit characterized by the stabilizer $G_{M}$ of one representative module $M$ from the orbit and by the associated $2$-cocycle $\vartheta_{M}$. In turn, each irreducible projective representation $\chb p \in \irr{G_{M}\vert\vartheta_{M}}$ corresponds to a primary field of the orbifold. We'll call the set of primaries originating from a particular $G$-orbit the block corresponding to that orbit. By the above, the primaries of the orbifold are partitioned into blocks, each corresponding to a $G$-orbit of twisted modules, and the primaries inside a given block correspond to the (projective) irreducible representations of the stabilizer of the orbit. We shall use the notation $\FC{\mathfrak{b}}$ to denote the $G$-orbit of twisted modules corresponding to a block $\mathfrak{b}$. Note that $\FA{\FC{\mathfrak{b}}}=\left[G : G_{M}\right]$ and $\FA{\mathfrak{b}}=\FA{\irr{G_{M}\vert\vartheta_{M}}}$ by the above, for any $M \in \FC{\mathfrak{b}}$.

If $M \in \FC{\mathfrak{b}}$ is a $g$-twisted module of ${\mathbb V}$ with $g \in G$ of order $n$, then the eigenvalues of the operator $nL_{0}$ are integrally spaced on $M$, i.e., the difference of two $L_{0}$ eigenvalues is an integer multiple of $\nicefrac{1}{n}$. We define the order of a block $\mathfrak{b}$ as the smallest positive integer $n$ such that the eigenvalues of the operator $nL_{0}$ are integrally spaced on any $M \in \FC{\mathfrak{b}}$. In particular, untwisted modules, are characterized by the property that their $L_{0}$ eigenvalues are integrally spaced, providing a simple criterion singling out untwisted modules from twisted ones. This also means that a block corresponding to a $G$-orbit of untwisted modules is characterized by the fact that the conformal weights of its members differ by integers, i.e., its order is $1$.

In particular, the vacuum module belongs to the untwisted sector and is left fixed by all elements of $G$, hence it forms in itself a $G$-orbit of length $1$ whose stabilizer is the whole of $G$, with trivial associated cocycle (because the twist group acts linearly on ${\mathbb V}$). It follows that the block corresponding to this vacuum module (the vacuum block $\mathfrak{b}_{\v}$) consists of primaries in one-to-one correspondence with the irreducible representations of~$G$, whose quantum dimensions are integers (since they equal the dimension of the corresponding irreducible representations), and whose conformal weights are integers too, because the conformal weight of the vacuum module is zero (recall that we consider unitary models).

\looseness=1 Since the vacuum acts as the identity of the fusion product, the fusion of an element of any given block~$\mathfrak{b}$ with an element of the vacuum block $\mathfrak{b_{\v}}$ will contain only primaries from block~$\mathfrak{b}$. This means that two primaries $p$ and~$q$ belong to the same block precisely when $N_{\alpha p}^{q} > 0$ for at least one element $\alpha \in \mathfrak{b_{\v}}$ of the vacuum block. In particular, the vacuum block generates a closed subring of the fusion ring of the orbifold, which is identical to the representation ring (Grothendieck ring) of the twist group, since the fusion product of any two of its members is the same as the tensor product of the corresponding irreducible representations.

\section{Orbifold deconstruction\label{sec:Orbifold-deconstruction}}

Armed with the above, we can now attack the problem of orbifold deconstruction, i.e., the identification of the original model and the twist group from the sole knowledge of data related to the orbifold. We start from a (unitary) conformal model for which we know the fusion coefficients $N_{pq}^{r}$, conformal weights $\cw p$, quantum dimensions $\qd p$ and chiral characters $\chi_{p} (\tau)$ of the primary fields, and we wish to identify it as a non-trivial orbifold of some other conformal model. Notice that one and the same model might have several deconstructions, reflecting the fact that the same model can be realized as an orbifold in many distinct ways; for example, if the twist group $G$ has a (non-trivial) normal subgroup $N \triangleleft G$, then a $G$-orbifold can be obtained as a $G/N$-orbifold of an $N$-orbifold. This indicates that there exists a whole hierarchy of deconstructions, whose most interesting members are the maximal ones leading to primitive models, i.e., models that cannot be obtained as a non-trivial orbifold of some other model, hence cannot be deconstructed any further.

The starting point of the deconstruction procedure is the observation, made at the end of the previous section, that the vacuum block of an orbifold has very special properties: it is a~set of primaries closed under the fusion product, and all its members have integer conformal weight and quantum dimension. For an arbitrary conformal model, we'll call a set of primaries with these properties a~twister. By the above, the vacuum block of an orbifold is a twister, and we assume in the sequel that all twisters arise as the vacuum block of a suitable orbifold realization of the model under study.\footnote{While we have no formal argument supporting this assumption, all available computational results corroborate it.} For each twister there is a different deconstruction of the model, and maximal deconstructions correspond to maximal twisters not contained in any other twister. The trivial twister (that contains the vacuum primary only) gives rise to a trivial deconstruction resulting in the original model.

Let's now consider the deconstruction with respect to a particular twister $\chg$, assumed to be non-trivial. The sole knowledge of the twister allows the determination of a host of important information about the twist group. For example, because the quantum dimensions of the ele\-ments of $\chg$ equal the dimensions of the corresponding irreducible representations of the twist group, and since the order of a group equals, by Burnside's famous theorem \cite{Isaacs,Serre}, the sum of the squared dimensions of its irreducible representations, it follows that the order of the twist group should equal the $\spr$
\begin{gather*}
\to=\sum_{\alpha\in\chg}\qd{\alpha}^{2}
\end{gather*}
of the twister. Along the same lines, the number of conjugacy classes of the twist group equals the size $\FA{\chg}$ of the twister, i.e., the number of its elements. Since the twister generates a subring of the fusion ring isomorphic to the representation ring of the twist group, and the knowledge of the representation ring determines the character table of the underlying group~\cite{Lux-Pahlings}, even the character table of the twist group and the size of its conjugacy classes may be computed from this information, and this is usually sufficient to identify the twist group up to isomorphism.\footnote{We note that, while the representation ring does not determine the group uniquely up to isomorphism (a~famous example being that of the dihedral group $\mathbb{D}_{4}$ of order $8$ and the group $\mathtt{Q}$ of unit quaternions, with identical representation rings), the braided structure of the module category of ${\mathbb V}$ allows one to determine the $\lambda$-ring structure of the representation ring, and in particular the powers of the conjugacy classes, leading to the possibility of identifying the twist group unambiguously.}

Once the twist group has been identified, the next problem is the determination of the primary fields of the original model and their most important attributes, like conformal weights, quantum dimensions and chiral characters. We know that these primaries originate in the untwisted sector of the orbifold, corresponding to untwisted modules organized into $G$-orbits, and to each such orbit corresponds a block of the orbifold, whose elements are in one-to-one correspondence with the projective irreducible representations of the stabilizer subgroup of the orbit (more precisely, of one module from it). These blocks, viewed as set of primaries of the orbifold, have the characteristic property that the primaries $p$ and $q$ belong to the same block precisely when $N_{\alpha p}^{q} > 0$ for at least one element $\alpha$ of the vacuum block. But the twister $\chg$ chosen for
deconstruction is precisely the vacuum block of the orbifold realization of our model, hence the blocks my be recovered from the knowledge of $\chg$ alone: these are the maximal sets $\mathfrak{b}$ of primaries such that $p,q \in \mathfrak{b}$ implies $N_{\alpha p}^{q} > 0$ for at least one element $\alpha \in \chg$.

Actually, there is no need to compute all the blocks for deconstruction, since only those corresponding to untwisted modules are needed for identifying the deconstructed model. But we know that blocks corresponding to a $G$-orbit of untwisted modules have order $1$, i.e., the conformal weights of their members can differ only by integers, providing a simple criterion to single them out.

Having identified the blocks in the untwisted sector, i.e., the $G$-orbits of (untwisted) modules of the deconstructed model, what remains is to determine the properties of the primary fields corresponding to these modules. Since they lie on the same $G$-orbit and are thus related by an automorphism of the relevant chiral algebra, they have the same conformal weights, quantum dimensions and chiral characters, so it is enough to determine these data for just one of them, but one should keep in mind that they correspond to different primaries of the deconstructed theory, hence their multiplicity (the length $\el{\mathfrak{b}}$ of the corresponding $G$-orbit) should be taken into account.

Since they are related by an element of $\operatorname{Aut}({\mathbb V})$, the modules $M$ belonging to the $G$-orbit $\FC{\mathfrak{b}}$ corresponding to a block $\mathfrak{b}$ all have the same conformal weight $\cw M=\min\set{\cw p}{p \in \mathfrak{b}}$, quantum dimension $\qd M$ and chiral character (trace function)
\begin{gather*}
\operatorname{Tr}_{M}\big\{ \mathtt{e}^{2\pi\mathtt{i}\tau(L_{0}-\nicefrac{c}{24})}\big\} =\sum_{p\in\mathfrak{b}}(\dim\chb p )\chi_{p}(\tau)=\frac{1}{\el{\mathfrak{b}}}\sum_{p\in\mathfrak{b}}\frac{\qd p}{\qd M}\chi_{p} (\tau), 
\end{gather*}
where $\el{\mathfrak{b}}$ denotes the length of the $G$-orbit $\FC{\mathfrak{b}}$. Note that this last result would provide an explicit form of the chiral characters should we know the quantum dimensions $\qd M$ and the orbit lengths $\el{\mathfrak{b}}$. Unfortunately, their determination could be tricky in general because the 2-cocycle $\vartheta_{M} \in Z^{2} (G_{M},\mathbb{C} )$ is usually non-trivial, and the irreducible representations $\chb p \in \irr{G_{M}\vert\vartheta_{M}}$ corresponding to the primaries $p \in \mathfrak{b}$ are not ordinary representations, but only projective ones. Nevertheless, exploiting general properties of projective representations it is possible to set up a~set of rules that allow their unambiguous computation in most instances. In particular, $\qd M$~is an algebraic integer $\geq 1$ such that the quantum dimensions $\qd p$ are all integer multiples
of $\el{\mathfrak{b}}\rami M\qd M$ for $p \in \mathfrak{b}$, where $\rami M$ denotes the multiplicative order of the cohomology class of~$\vartheta_{M}$ (which is obviously the same for all modules $M \in \FC{\mathfrak{b}}$).
These observations are usually sufficient to pin down the precise values of~$\qd M$ and~$\el{\mathfrak{b}}$.

Finally, to finish the identification of the deconstructed model, one needs to find out its fusion rules. This can be accomplished by assigning to each block $\mathfrak{b}$ of the untwisted sector a~block-fusion matrix $\fm (\mathfrak{b})$ with matrix elements
\begin{gather*}
\fm (\mathfrak{b})_{q}^{r}=\frac{1}{\to}\sum_{p\in\mathfrak{b}}\frac{\qd p}{\qd M}N_{pq}^{r}.
\end{gather*}
Such matrices\emph{ }form a ring, i.e., the product of any two of them may be expressed as a sum
\begin{gather*}
\fm (\mathfrak{a})\fm (\mathfrak{b})=\sum_{\mathfrak{c}}\mathfrak{\bfm_{ab}^{c}}\fm (\mathfrak{c})
\end{gather*}
over the untwisted blocks, where the block-fusion coefficients $\mathfrak{\bfm_{ab}^{c}} \in \mathbb{Z_{+}}$ are given by the sums
\begin{gather}
\mathfrak{\bfm_{ab}^{c}}=\sum_{A\in\FC{\mathfrak{a}},\,B\in\FC{\mathfrak{b}}}N_{AB}^{C}\label{eq:blockfusion}
\end{gather}
for any representative module $C \in \FC{\mathfrak{c}}$. Note that these block-fusion coefficients come near to provide the fusion rules of the deconstructed model, but for one thing: they do not describe the fusion of the individual modules, but only that of the direct sum of the modules contained in the orbits corresponding to the individual blocks. This shouldn't come as a surprise since, after all, orbifolding amounts to identifying the modules related by a symmetry, so that their individual properties are lost in the process, except for those (like conformal weights and quantum dimensions) which are the same for all modules on the same orbit. Nevertheless, this aggregated version of the fusion rules, supplemented by the knowledge of the conformal weights, quantum dimensions and chiral characters is enough in many examples to identify uniquely the deconstructed model, at least
in those amenable to direct computations.\footnote{We should also point out that, since the block-fusion matrices depend explicitly on the quantum dimensions~$\qd M$ but not on the orbit lengths $\el{\mathfrak{b}}$, the integrality of the block-fusion coefficients $\mathfrak{\bfm_{ab}^{c}}$ provides an extra condition for determining the precise value of the dimensions in case they are not fixed by the previous considerations.}

\section[Example: the 3-fold symmetric product of the moonshine module]{Example: the 3-fold symmetric product\\ of the moonshine module}
\global\long\def\moon{\mathbb{V}^{\natural}}%
\global\long\def\vac{\mathtt{0}}%
\newcommandx\mpru[2][usedefault, addprefix=\global, 1=]{\boldsymbol{#2}_{{\scriptscriptstyle #1}}}%
\global\long\def\mprs#1{\boldsymbol{\upsigma}_{{\scriptscriptstyle #1}}}%
\global\long\def\mprc#1{\boldsymbol{\uprho}_{#1}}%

Consider the self-dual $c=72$ conformal model made up of three identical copies of the moonshine module $\moon$. Clearly, any permutation of these identical copies is a symmetry, hence one may orbifold this model with respect to the symmetric group $\sn 3$ of degree $3$, resulting in the $3$-fold symmetric product $\moon\wr\mathbb{S}_{3}$ of the moonshine module \cite{Bantay2003a, SymProd}. The properties of this symmetric product are well understood, since it is by construction a permutation orbifold \cite{Bantay2002}, and at the same time a~holomorphic orbifold \cite{DV3}, because the moonshine module is self-dual.

$\moon\wr\mathbb{S}_{3}$ is known to have $8$ different primaries \cite{Bantay2002}, whose most important properties are summarized in the following table (with $\zeta=\exi[][2]3$ denoting a primitive third root of unity).

\global\long\def\sro{\boldsymbol{\text{\ensuremath{\varSigma}}}}%
\global\long\def\su{\boldsymbol{\varXi}}%

\begin{table}[h!]\centering
\begin{tabular}[t]{|l||c|c|c|}
\hline
label & conformal weight & dimension & character\tabularnewline
\hline
\hline
\multirow{2}{*}{$\mpru[+]{~~1}$} &\multirow{2}{*}{$0$} & \multirow{2}{*}{$1$} & \multirow{2}{*}{${\displaystyle \frac{1}{6}}\left\{ J (\tau)^{3}+3J (\tau)J \left(2\tau\right)+2J \left(3\tau\right)\right\} $}\tabularnewline
 & & & \tabularnewline
\hline
\multirow{2}{*}{$~~\mpru[-]1$} &\multirow{2}{*}{$4$} & \multirow{2}{*}{$1$} & \multirow{2}{*}{${\displaystyle \frac{1}{6}}\left\{ J (\tau)^{3}-3J (\tau)J \left(2\tau\right)+2J \left(3\tau\right)\right\} $}\tabularnewline
 & & & \tabularnewline
\hline
 \multirow{2}{*}{$~~\mpru 2$} & \multirow{2}{*}{$2$} &\multirow{2}{*}{$2$} & \multirow{2}{*}{${\displaystyle \frac{1}{3}}\left\{ J (\tau)^{3}-J \left(3\tau\right)\right\} $}\tabularnewline
 & & & \tabularnewline
\hline
\multirow{2}{*}{$~~\mprc 0$} & \multirow{2}{*}{${\displaystyle \frac{8}{3}}$} & \multirow{2}{*}{$2$} & \multirow{2}{*}{${\displaystyle \frac{1}{3}}\left\{ J \left(\frac{\tau}{3}\right)+\zeta J \left(\frac{\tau+1}{3}\right)+\overline{\zeta}J \left(\frac{\tau+2}{3}\right)\right\} $}\tabularnewline
 & & & \tabularnewline
\hline
\multirow{2}{*}{$~~\mprc 1$} & \multirow{2}{*}{$4$} & \multirow{2}{*}{$2$} & \multirow{2}{*}{${\displaystyle \frac{1}{3}}\left\{ J \left(\frac{\tau}{3}\right)+J \left(\frac{\tau+1}{3}\right)+J \left(\frac{\tau+2}{3}\right)\right\} $}\tabularnewline
 & & & \tabularnewline
\hline
\multirow{2}{*}{$~~\mprc 2$} & \multirow{2}{*}{${\displaystyle \frac{10}{3}}$} & \multirow{2}{*}{$2$} & \multirow{2}{*}{${\displaystyle \frac{1}{3}}\left\{ J \left(\frac{\tau}{3}\right)+\overline{\zeta}J \left(\frac{\tau+1}{3}\right)+\zeta J \left(\frac{\tau+2}{3}\right)\right\} $}\tabularnewline
 & & & \tabularnewline
\hline
 \multirow{2}{*}{$~~\mprs +$}& \multirow{2}{*}{${\displaystyle \frac{3}{2}}$} & \multirow{2}{*}{$3$} & \multirow{2}{*}{${\displaystyle \frac{1}{2}}J (\tau)\left\{ J \left(\frac{\tau}{2}\right)+J \left(\frac{\tau+1}{2}\right)\right\} $}\tabularnewline
 & & & \tabularnewline
\hline
 \multirow{2}{*}{$~~\mprs -$}& \multirow{2}{*}{$3$}& \multirow{2}{*}{$3$} & \multirow{2}{*}{${\displaystyle \frac{1}{2}}J (\tau)\left\{ J \left(\frac{\tau}{2}\right)-J \left(\frac{\tau+1}{2}\right)\right\} $}\tabularnewline
 & & &\tabularnewline
\hline
\end{tabular}
\caption{Primaries of $\moon\wr\mathbb{S}_{3}$.}
\end{table}

It follows that the only primaries that can belong to a twister are among $\mpru[+]1$, $\mpru[-]1$, $\mpru 2$, $\mprc 1$ and~$\mprs -$, since only these have simultaneously integer quantum dimension and conformal weight.

With the notations $\su=\mpru[+]1 \oplus \mpru[-]1$ and $\sro=\mpru 2 \oplus \mprc 0 \oplus \mprc 1 \oplus \mprc 2$, the fusion rules read

\begin{table}[h!]\centering
\begin{tabular}[t]{|l||c|c|c|c|c|c|c|c|}
\hline
 & $\mpru[+]1$ & $\mpru[-]1$ & $~~\mpru 2$ & $~~\mprc 0$ & $~~\mprc 1$ & $~~\mprc 2$ & $~~\mprs +$ & $~~\mprs -$\tabularnewline
\hline
\hline
$\mpru[+]1$ & $\mpru[+]1$ & $\mpru[-]1$ & \multirow{1}{*}{$~~\mpru 2$} & $~~\mprc 0$ & $~~\mprc 1$ & $~~\mprc 2$ & $~~\mprs +$ & $~~\mprs -$\tabularnewline
\hline
$\mpru[-]1$ & $\mpru[-]1$ & $\mpru[+]1$ & \multirow{1}{*}{$~~\mpru 2$} & $~~\mprc 0$ & $~~\mprc 1$ & $~~\mprc 2$ & $\mprs -$ & $\mprs +$\tabularnewline
\hline
$\mpru 2$ & $\mpru 2$ & $\mpru 2$ & \multirow{1}{*}{$\mpru 2 \oplus \su$} & $\mprc 1 \oplus \mprc 2$ & $\mprc 0 \oplus \mprc 2$ & $\mprc 0 \oplus \mprc 1$ & $\mprs + \oplus \mprs -$ & $\mprs + \oplus \mprs -$\tabularnewline
\hline
$\mprc 0$ & $\mprc 0$ & $\mprc 0$ & $\mprc 1 \oplus \mprc 2$ & $\mprc 0 \oplus \su$ & $\mpru 2 \oplus \mprc 2$ & $\mpru 2 \oplus \mprc 1$ & $\mprs + \oplus \mprs -$ & $\mprs + \oplus \mprs -$\tabularnewline
\hline
$\mprc 1$ & $\mprc 1$ & $\mprc 1$ & $\mprc 0 \oplus \mprc 2$ & $\mpru 2 \oplus \mprc 2$ & $\mprc 1 \oplus \su$ & $\mpru 2 \oplus \mprc 0$ & $\mprs + \oplus \mprs -$ & $\mprs + \oplus \mprs -$\tabularnewline
\hline
$\mprc 2$ & $\mprc 2$ & $\mprc 2$ & $\mprc 0 \oplus \mprc 1$ & $\mpru 2 \oplus \mprc 1$ & $\mpru 2 \oplus \mprc 0$ & $\mprc 2 \oplus \su$ & $\mprs + \oplus \mprs -$ & $\mprs + \oplus \mprs -$\tabularnewline
\hline
$\mprs +$ & $\mprs +$ & $\mprs -$ & $\mprs + \oplus \mprs -$ & $\mprs + \oplus \mprs -$ & $\mprs + \oplus \mprs -$ & $\mprs + \oplus \mprs -$ & $\mpru[+]1 \oplus \sro$ & $\mpru[-]1 \oplus \sro$\tabularnewline
\hline
$\mprs -$ & $\mprs -$ & $\mprs +$ & $\mprs + \oplus \mprs -$ & $\mprs + \oplus \mprs -$ & $\mprs + \oplus \mprs -$ & $\mprs + \oplus \mprs -$ & $\mpru[-]1 \oplus \sro$ & $\mpru[+]1 \oplus \sro$\tabularnewline
\hline
\end{tabular}
\caption{Fusion rules of $\moon \wr \mathbb{S}_{3}$.}
\end{table}

It follows that there are precisely three non-trivial twisters: the maximal twisters $ \{ \mpru[+]1,\mpru[-]1,\mpru 2 \} $ and $\{ \mpru[+]1,\mpru[-]1,\mprc 1\} $, both of $\spr$ $6$ and size~$3$, and their intersection $\{ \mpru[+]1,\mpru[-]1\} $. We note that the existence of two maximal twisters is related to the  automorphism of the fusion rules that exchanges the primaries $\mpru 2$ and $\mprc 1$ while leaving all other primaries invariant. Let's take a closer look at the different deconstructions corresponding to these cases.\footnote{An important feature of this example that simplifies tremendously the analysis is that, because the relevant groups are pretty small, all the cocycles are trivial, so one does only encounter ordinary representations during the procedure.}

The twister $\{ \mpru[+]1,\mpru[-]1\} $ consists of two simple currents corresponding to the one-dimensional representations of $\sn 3$, hence the deconstructed model will be a simple current extension of $\moon \wr \sn 3$ that could be determined alternatively using the techniques of~\cite{Fuchs1996a}. Obviously, the twist group is isomorphic to~$\mathbb{Z}_{2}$. There are~$6$ different blocks, whose properties can be read off the following table (the multiplicity of a~block~$\mathfrak{b}$ is the length $\FA{\FC{\mathfrak{b}}}$ of the associated orbit, cf.\ Section~\ref{sec:Orbifold-deconstruction}).

\begin{table}[h!]\centering
\begin{tabular}{|c||c|c|c|c|}
\hline
block & order & multiplicity & dimension & trace function\tabularnewline
\hline
\hline
$\left\{ \mpru[+]1,\mpru[-]1\right\} $ & $1$ & $1$ & $1$ & ${\displaystyle {\textstyle \frac{1}{3}}}\left\{ J (\tau)^{3}+2J \left(3\tau\right)\right\} $\tabularnewline
\hline
$\left\{ \mpru 2\right\} $ & $1$ & $2$ & $1$ & ${\textstyle \frac{1}{3}}\left\{ J (\tau)^{3}-J \left(3\tau\right)\right\} $\tabularnewline
\hline
$\left\{ \mprc 0\right\} $ & $1$ & $2$ & $1$ & ${\displaystyle {\textstyle \frac{1}{3}}}\left\{ J \left(\frac{\tau}{3}\right)+\zeta J \left(\frac{\tau+1}{3}\right)+\overline{\zeta}J \left(\frac{\tau+2}{3}\right)\right\} $\tabularnewline
\hline
$\left\{ \mprc 1\right\} $ & $1$ & $ 2$ & $1$ & ${\displaystyle {\textstyle \frac{1}{3}}}\left\{ J \left(\frac{\tau}{3}\right)+J \left(\frac{\tau+1}{3}\right)+J \left(\frac{\tau+2}{3}\right)\right\} $\tabularnewline
\hline
$\left\{ \mprc 2\right\} $ & $1$ & $2$ & $1$ & ${\displaystyle {\textstyle \frac{1}{3}}}\left\{ J \left(\frac{\tau}{3}\right)+\overline{\zeta}J \left(\frac{\tau+1}{3}\right)+\zeta J \left(\frac{\tau+2}{3}\right)\right\} $\tabularnewline
\hline
$\left\{ \mprs +,\mprs -\right\} $ & $2$ & $1$ & $3$ & $J (\tau)J \left(\frac{\tau}{2}\right)$\tabularnewline
\hline
\end{tabular}
\end{table}

The untwisted sector consists of $9$ modules arranged into $4$ orbits of length $2$ and one fixed-point (the vacuum block). Inspecting their conformal weights and trace functions, one recognizes that the deconstructed model is nothing but the permutation orbifold $\moon \wr \mathbb{A}_{3}$ of the moonshine module by the alternating group $\mathbb{A}_{3}$ of degree $3$ (the commutator subgroup of $\sn 3$); this is further corroborated by the fusion rules computed using equation~\eqref{eq:blockfusion}. We note that there is a unique $\mathbb{Z}_{2}$-twisted module in this case, corresponding to the block of order $2$.

For the twister $\{ \mpru[+]1,\mpru[-]1,\mpru 2\} $, the fusion rules are

\begin{table}[h!]\centering
\begin{tabular}{|l||c|c|c|}
\cline{2-4}
\multicolumn{1}{l|}{} & $\mpru[+]1$ & $\mpru[-]1$ & $\mpru 2$\tabularnewline
\hline
\hline
$\mpru[+]1$ & $\mpru[+]1$ & $\mpru[-]1$ & $\mpru 2$\tabularnewline
\hline
$\mpru[-]1$ & $\mpru[-]1$ & $\mpru[+]1$ & $\mpru 2$\tabularnewline
\hline
$\mpru 2$ & $\mpru 2$ & $\mpru 2$ & $\mpru[+]1 \oplus \mpru[-]1 \oplus \mpru 2$\tabularnewline
\hline
\end{tabular}
 \end{table}

\noindent leading to the following character table for the twist group

\begin{table}[h!]\centering
\begin{tabular}{|l||r|r|r|}
\cline{2-4}
\multicolumn{1}{l|}{} & $\tcl$ & $\boldsymbol{\mathfrak{2}}$ & $\boldsymbol{\mathfrak{3}}$\tabularnewline
\hline
\hline
$\mpru[+]1$ & $1$ & $1$ & $1$\tabularnewline
\hline
$\mpru[-]1$ & $1$ & $-1$ & $1$\tabularnewline
\hline
$\mpru 2$ & $2$ & $0$ & $-1$\tabularnewline
\hline
\end{tabular}
 \end{table}

\noindent
from which can one infer that the twist group is isomorphic to $\mathbb{S}_{3}$ in this case. Actually, this already follows from the fact that it is a group of order $\to=6$ with $\FA{\chg}=3$ different conjugacy classes, and all such groups are isomorphic to $\sn 3$.

There are $2$ blocks besides the vacuum block, and the following table summarizes their most important properties.

\begin{table}[h!]\centering
\begin{tabular}{|c||c|c|c|c|}
\hline
block & order & multiplicity & dimension & trace function\tabularnewline
\hline
\hline
$\left\{ \mpru[+]1,\mpru[-]1,\mpru 2\right\} $ & $1$ & $1$ & $1$ & $J (\tau)^{3}$\tabularnewline
\hline
$\left\{ \mprs +,\mprs -\right\} $ & $2$ & $3$ & $1$ & $J (\tau)J \left(\frac{\tau}{2}\right)$\tabularnewline
\hline
$\left\{ \mprc 0,\mprc 1,\mprc 2\right\} $ & $3$ & $2$ & $1$ & $J \left(\frac{\tau}{3}\right)$\tabularnewline
\hline
\end{tabular}
 \end{table}

Since there is just one block in the untwisted sector (i.e., of order~$1$), it follows that the deconstructed model is self-dual, with trace function equal to $J (\tau)^{3}$. From this we can conclude that the deconstructed model is made up from $3$ identical copies of the moonshine module $\moon$.

For the twister $ \{ \mpru[+]1,\mpru[-]1,\mprc 1 \} $, the  situation is pretty similar to the previous one after one performs the exchange $\mpru 2\leftrightarrow\mprc 1$. The fusion rules read

\begin{center}
\begin{tabular}{|l||c|c|c|}
\cline{2-4}
\multicolumn{1}{l|}{} & $\mpru[+]1$ & $\mpru[-]1$ & $\mprc 1$\tabularnewline
\hline
\hline
$\mpru[+]1$ & $\mpru[+]1$ & $\mpru[-]1$ & $\mprc 1$\tabularnewline
\hline
$\mpru[-]1$ & $\mpru[-]1$ & $\mpru[+]1$ & $\mprc 1$\tabularnewline
\hline
$\mprc 1$ & $\mprc 1$ & $\mprc 1$ & $\mpru[+]1 \oplus \mpru[-]1 \oplus \mprc 1$\tabularnewline
\hline
\end{tabular}
 \end{center}

\noindent leading once again to the character table

\begin{center}
\begin{tabular}{|l||r|r|r|}
\cline{2-4}
\multicolumn{1}{l|}{} & $\tcl$ & $\boldsymbol{\mathfrak{2}}$ & $\boldsymbol{\mathfrak{3}}$\tabularnewline
\hline
\hline
$\mpru[+]1$ & $1$ & $1$ & $1$\tabularnewline
\hline
$\mpru[-]1$ & $1$ & $-1$ & $1$\tabularnewline
\hline
$\mprc 1$ & $2$ & $0$ & $-1$\tabularnewline
\hline
\end{tabular}\end{center}

\noindent
showing that the twist group is isomorphic to $\mathbb{S}_{3}$ in this case too. Once again, there are $3$ blocks: the vacuum block $\{ \mpru[+]1,\mpru[-]1,\mprc 1\} $, and the blocks $\{\boldsymbol{\mprs +},\boldsymbol{\mprs -}\}$ and $\{ \mprc 0,\mpru 2,\mprc 2\} $ of respective orders~$2$ and~$3$. These share the properties of their counterparts for the twister $\{ \mpru[+]1,\mpru[-]1,\mpru 2\} $, except for their trace functions, which read in this case

\begin{table}[h!]\centering
\begin{tabular}{|c||c|}
\hline
block & trace function\tabularnewline
\hline
\hline
$\left\{ \mpru[+]1,\mpru[-]1,\mprc 1\right\} $ & $J (\tau)^{3}-2c_{1}J (\tau)-2c_{2}$\tabularnewline
\hline
$\left\{ \mprs +,\mprs -\right\} $ & $J (\tau)J \left(\frac{\tau}{2}\right)$\tabularnewline
\hline
$\left\{ \mprc 0,\mpru 2,\mprc 2\right\} $ & $ J \left(\frac{\tau}{3}\right)+c_{1}J (\tau)+c_{2}$\tabularnewline
\hline
\end{tabular}
 \end{table}

\noindent upon taking into account the replication identity\footnote{\noindent This illustrates the fact that the existence of different maximal deconstructions of a permutation orbifold is related to the existence of non-trivial replication identities~\cite{Bantay2016}.}
\begin{gather*}
J \left(\frac{\tau}{3}\right)+J \left(\frac{\tau+1}{3}\right)+J \left(\frac{\tau+2}{3}\right)+J \left(3\tau\right)=J (\tau)^{3}-3c_{1}J (\tau)-3c_{2}
\end{gather*}
for the modular invariant function $J (\tau)$, where $c_{1}=196884$ and $c_{2}=21493760$.

Note that this second deconstruction differs markedly from the previous one, although their twist groups are isomorphic. In particular, since the polynomial $P(x)=x^{3}-2c_{1}x-2c_{2}$ has $3$ different roots, it follows that the action of the twist group is not a permutation action in this case. This exemplifies that one and the same conformal model (in our case $\moon\wr\mathbb{S}_{3}$) may be obtained as an orbifold of quite different models, with different (possibly non-isomorphic) twist groups.

\section{Summary and outlook}

As described above, there is an effective algorithmic procedure for orbifold deconstruction, i.e., the realization of a given conformal model as an orbifold of some other model. Different deconstructions correspond to different twisters of the model under study, where non-maximal deconstructions lead to models that can be further deconstructed themselves, corresponding to orbifolding in stages. Deconstructions corresponding to maximal twisters (i.e., not contained in any other twister) are the ones that may lead to primitive models that cannot be realized as non-trivial orbifolds.\footnote{It may happen that orbifolding by stages breaks down, i.e., the orbifold of an orbifold is not an orbifold of the original model, in which case the corresponding maximal deconstruction is itself a non-trivial orbifold. This phenomenon shows up, as put forward by one of the referees, in the 16-fold tensor power of the Ising model, which has a maximal deconstruction to the ${\rm SO}(16)$ Wess--Zumino model at level 1, itself a $\mathbb{Z}_{2}$-orbifold of the $E_{8}$ Wess--Zumino model of the same level.} The deconstruction algorithm provides us with a more-or-less unique identification of the twist group and of the original model, up to some ambiguities related to the projective realization of the twist group and to the individual fusion rules of the untwisted modules. It should be emphasized that this ambiguity is a finite one, which means that, as a last resort, one can in principle go through a tedious case-by-case analysis of the allowed possibilities to find out the correct one.

Besides its intrinsic interest, what could be the benefits of the deconstruction procedure in the study of 2D CFT? There is one such benefit that is more or less obvious: if we can identify both the deconstructed model and the twist group unambiguously, then we get an example of an orbifold construction where the result, being the input of the deconstruction procedure, is known right from the start, making possible a thorough investigation of the orbifolding process. Actually, the procedure can be extended so as to give precise results not only about the modules in the untwisted sector (the ones that are of interest for identifying the deconstructed model), but also about the structure of all of the twisted modules. This is particularly important, since the structure of $g$-twisted modules does only depend on the conjugacy class of $g$ in $\operatorname{Aut}({\mathbb V})$, hence the deconstruction of a~$G$-orbifold gives valuable information about the structure of all orbifolds (of the deconstructed model) whose twist group contains elements that are conjugate in $\operatorname{Aut}({\mathbb V})$ to some element of~$G$.

Finally, let us note that the above ideas might be used in attempts to classify rational conformal models. For one thing, the classification problem can be reduced to that of primitive models (the ones that don't have nontrivial twisters), since all others are orbifolds of these, and the latter can be classified by group theoretic means. On the other hand, even primitive models can be grouped together if they arise from maximal deconstructions of one and the same conformal model, i.e., if some of their orbifolds are identical. An interesting problem is to develop simple criteria testing whether two primitive models are related in this way, and to understand what kind of common structures are responsible for such behavior.

\pdfbookmark[1]{References}{ref}
\LastPageEnding

\end{document}